\documentclass[aps,twocolumn,showpacs]{revtex4}
\usepackage{amsmath}
\usepackage{epsfig}

\newcommand{\xc} {exchange-correlation}

\newcommand{\vecr} {{\mathbf r}}
\newcommand{\veck} {{\mathbf k}}
\newcommand{\vecq} {{\mathbf q}}
\newcommand{\vecG} {{\mathbf G}}
\newcommand{\vecp} {{\mathbf p}}
\newcommand{\vece} {{\mathbf e}}

\newcommand{\Exc} {E_{xc}}

\newcommand{\vKS} {v_{KS}}
\newcommand{\vext} {v_{ext}}

\newcommand{\FH} {F_H}
\newcommand{\Fxc} {F_{xc}}

\newcommand{\phiik} {\phi_{i \veck}}
\newcommand{\evk} {\epsilon_{v \veck}}

\newcommand{\eik} {\epsilon_{i \veck}}

\newcommand{\qzero} {q \rightarrow 0}

\begin{document}

\title{Optical excitations of Si by time-dependent density-functional
theory based on the exact-exchange Kohn-Sham band structure}

\author{Yong-Hoon Kim$^1$, Martin St{\"a}dele$^2$, and
		Andreas G{\"o}rling$^1$}

\address{\hspace{-0.5cm}$^1$ Lehrstuhl f{\"u}r Theoretische Chemie,
	Technische Universit{\"a}t M{\"u}nchen,
	D-85748 Garching, Germany}

\address{$^2$ Infineon Technologies AG, CPR ND,
	Otto-Hahn-Ring 6, D-81730 Munich, Germany}



\begin{abstract}
We calculate the imaginary part of the frequency-dependent dielectric
function of bulk silicon by applying time-dependent density-functional
theory based on the exact-exchange (EXX) Kohn-Sham (KS) band structure
and the adiabatic local-density approximation (ALDA) kernel. The
position of the $E_2$ absorption peak calculated with the EXX band
structure at the independent-particle level is in excellent agreement
with experiments, which demonstrates the good quality of EXX `KS
quasiparticles'. The excitonic $E_1$ peak that is missing at the
independent-particle level remains absent if two-particle interaction
effects are taken into account within the time-dependent LDA,
demonstrating the incapability of the ALDA kernel to describe
excitonic effects.
\end{abstract}


\maketitle

\section{Introduction}
\label{sec:introduction}

While the Kohn-Sham (KS) density-functional theory
(DFT)~\cite{Kohn99,Jones89,Koch00} has become the dominant method of
first-principles investigations of energetic properties of solids, its
scope in the treatment of electronic excitations has been rather
limited due to deficiencies of the conventional local density
approximation (LDA) and its semilocal extensions. One well-known case
is the band gaps calculated within the LDA which are substantially
smaller than experimental values~\cite{Jones89}. Significant
advancement has been made during the past decades in devising improved
\xc\ functionals starting from the LDA~\cite{Koch00}, but these
semilocal approximations still fail to correct the so-called `band-gap
problem'. Thus for the accurate treatment of band structures one at
present usually leaves the domain of DFT and employs a many-body method
based on the GW approximation (GWA) to quasiparticle
energies~\cite{Hybertsen85,Aulbur00}.

For the case of optical excitations in which bound exciton states can
play a significant role, it is further required to go beyond effective
one-particle approaches.  In principle, this can be done within the
time-dependent (TD) DFT~\cite{Gross96,Casida95} or by solving the
Bethe-Salpeter equation (BSE) for two-body Green's
functions~\cite{Hanke79}.  However, the lack of a proper \xc\
functional again remains as the hindrance to the TDDFT route, and the
GWA-BSE currently represents the state-of-the-art method for such
purpose~\cite{Albrecht98,Benedict98,Rohlfing98}.

In this regard recent numerical realization of the exact-exchange
(EXX) KS method for solids~\cite{Bylander95,Gorling96,Stadele97} and
molecules~\cite{Grabo97,KLI99,Gorling99,Ivanov99,Veseth01,DellaSalla01}
provides an interesting opportunity: According to theoretical and
numerical evidence concerning the physical meaningfulness of the KS
potential and KS eigenvalues~\cite{Gorling96b,Umrigar98,Al-Sharif98},
realistic band structures for semiconductors and eigenvalue spectra of
molecules obtained within the EXX suggests a promising basis for the
study of electronic excited states within DFT. In this work, we
consider this point further by calculating the absorption spectrum of
bulk silicon within TDDFT based on the EXX band structure and the
adiabatic LDA (ALDA) \xc\ kernel. Silicon here serves as a
representative semiconductor system, whose first absorption peak, the
$E_1$ peak, is known to be strongly excitonic in its character. The
large excitonic contributions to the $E_1$ peak has been proposed
theoretically and experimentally, and a semiempirical calculation
confirming this point has been performed two decades
ago~\cite{Hanke79}. However, truly first-principles calculations
appeared only recently based on the
GWA-BSE~\cite{Albrecht98,Benedict98,Rohlfing98}. A different approach
to investigate optical spectra within a DFT framework was recently
explored in Ref.~\onlinecite{deBoeij01}, where the domain of
conventional DFT was left and time-dependent current DFT was
invoked. As a first step towards an accurate computation of optical
spectra within TDDFT, we will examine the performance of the EXX-TDLDA
approach in this paper.

\section{Theory}
\label{sec:theory}

To obtain the absorption spectrum of solids, we compute the imaginary
part of the frequency-dependent macroscopic dielectric function
$\epsilon(\omega) = \epsilon_1(\omega) + i \epsilon_2(\omega)$ defined
as~\cite{Pick70,Hybertsen87}
\begin{equation}
\label{eq:df-mac}
 \epsilon(\omega) =
	\Bigl[ \lim_{\qzero} \epsilon^{-1}(\vecq+\vecG,\vecq+\vecG';\omega)
		\bigl|_{\vecG=\vecG'=0}
	\Bigr]^{-1}
\end{equation}
where
\begin{equation}
\label{eq:df-inv}
\begin{split}
 \epsilon^{-1}(\vecq+\vecG,\vecq+\vecG';\omega) & =
	\delta_{\vecG,\vecG'} +
	\FH(\vecq + \vecG) \\
 & \times \chi(\vecq+\vecG,\vecq+\vecG';\omega),
\end{split}
\end{equation}
with $\FH(\vecq + \vecG) \equiv 4 \pi / |\vecq + \vecG|^2$ (Hartree
atomic units are used throughout the paper. The symbols $\vecq+\vecG$
refer to plane waves with wave vectors $\vecq+\vecG$,
$e^{i(\vecq+\vecG) \cdot \vecr}$, i.e., all formulas are given in the
plane wave representation). The linear response function $\chi$ that
describes the full response of the first-order density change $\delta
n$ for the given external perturbation $\delta \vext$,
\begin{equation}
\label{eq:dn-chi}
 \delta n(\vecq+\vecG;\omega) =
	\sum_{\vecG'} \ \chi(\vecq+\vecG,\vecq+\vecG';\omega) \
	\delta \vext(\vecq+\vecG';\omega),
\end{equation}
can be computed either via the BSE or TDDFT.  Within TDDFT the linear
response function $\chi$ is given in matrix representation
as~\cite{Gross96,Casida95}
\begin{equation}
\label{eq:chi-chi0}
 \chi(\vecq;\omega) =
	\Bigl[ 1 - \chi_0(\vecq;\omega) \{\FH(\vecq)+\Fxc(\vecq;\omega)\} \Bigr]^{-1}
	\chi_0(\vecq;\omega).
\end{equation}
where $\chi_0$ is the noninteracting KS linear response matrix. It is
given in terms of the occupied and unoccupied KS orbitals
$\phi_{v\veck}$ and $\phi_{c\veck}$ and their eigenvalues
$\epsilon_{v\veck}$ and $\epsilon_{c\veck}$ through
\begin{equation}
\label{eq:chi0}
\begin{split}
 & \chi_0 (\vecq+\vecG,\vecq+\vecG';\omega) = 
  \frac{2}{N \Omega} \\
 & \times \sum_{v,c,\veck} 
	\Biggl[ \frac{ \langle v, \veck| e^{-i(\vecq+\vecG) \cdot \vecr}
			| c,\veck+\vecq \rangle
	\langle c, \veck+\vecq| e^{i(\vecq+\vecG') \cdot \vecr'}
			|v,\veck \rangle }
	{ \evk - \epsilon_{c({\veck+\vecq})} + \omega + i \delta } \\
 & + \frac{ \langle c, \veck| e^{-i(\vecq+\vecG) \cdot \vecr }
		   |v,\veck+\vecq \rangle
	\langle v, \veck+\vecq| e^{i(\vecq+\vecG') \cdot \vecr'}
		   |c,\veck \rangle }
	{ \evk - \epsilon_{c({\veck+\vecq})} - \omega - i \delta }
	\Biggr],
\end{split}
\end{equation}
where $N \Omega$ is the crystal volume. In Eq. (\ref{eq:chi-chi0})
$\Fxc$ denotes the \xc\ kernel, the frequency-dependent functional
derivative of the \xc\ potential.

Thus, the macroscopic dielectric function $\epsilon$ can be computed by
first calculating $\epsilon^{-1}$ in the reciprocal-space using Eqs.
(\ref{eq:chi-chi0}) and (\ref{eq:chi0}), and next taking the inverse of
the $\qzero$ limit of its $\vecG=\vecG'=0$ element [Eq.
(\ref{eq:df-mac})]. However, the ALDA kernel $\Fxc^{ALDA} = \delta^2
\Exc^{LDA} / (\delta n \ \delta n')$ is local in real space and
frequency-independent, which results in a reciprocal-space
representation for $\Fxc$ that is independent of $q$ and $\omega$,
$\Fxc^{ALDA}(\vecq+\vecG,\vecq+\vecG';\omega) =
\Fxc^{ALDA}(\vecG-\vecG')$. This {\em incorrect} $\qzero$ behavior of
the ALDA kernel then simplifies the calculation of $\epsilon(\omega)$:
Writing $\chi$, $\FH$, and $\Fxc$ (from now on the superscript
``ALDA'' on the kernel is suppressed for notational simplicity) for
$\qzero$ in block form in terms of the ``head" [$\vecG=\vecG'=0$
(00)], ``wings" [$\vecG=0;\vecG' \neq 0$ (01) and $\vecG \neq
0;\vecG'=0$ (10)], and the ``body" [$\vecG \neq 0;\vecG'
\neq 0$ (11)] as
\begin{equation}
\begin{split}
 \chi_0(q;\omega) & =
	Q \begin{pmatrix}
		\chi_0^{00}(\omega) & \chi_0^{01}(\omega) \\
		\chi_0^{10}(\omega) & \chi_0^{11}(\omega)
	\end{pmatrix} Q ; \\
 \FH(q) & = Q^{-1} \begin{pmatrix}
		4 \pi & 0 \\
		0 & \FH^{11} \end{pmatrix} Q^{-1} ; \\
 \Fxc(q) & = Q^{-1} \begin{pmatrix}
		0 & 0 \\
		0 & \Fxc^{11} \end{pmatrix} Q^{-1}
\end{split}
\end{equation}
where $Q^{00} = q; Q^{01}=Q^{10}=0; Q^{11}=\delta_{\vecG,\vecG'}$, and
performing matrix multiplications explicitly (with
$\vecq$-singularities treated properly at each stage) yields a $3
\times 3$ tensor~\cite{Pick70,Hybertsen87},
\begin{equation}
\label{eq:DF-LDA}
\begin{split}
 \epsilon(\omega) = 1
	- 4 \pi & \chi_0^{00}(\omega)
	- 4 \pi \chi_0^{01}(\omega) (\FH^{11} + \Fxc^{11}) \\
	 \times &  [ 1 - \chi_0^{11}(\omega) (\FH^{11} + \Fxc^{11}) ]^{-1}
	\chi_0^{10}(\omega),
\end{split}
\end{equation}
with which matrix multiplications including head and wings with
explicit $\vecq$-dependence are avoided.  Eq. (\ref{eq:DF-LDA}) simply
represents the independent-particle-level absorption spectrum if
$\FH^{11}$ and $\Fxc^{11}$ are ignored altogether, while it is reverted
to the so-called RPA dielectric function formula if only $\Fxc^{11}$ is
neglected. For cubic systems, such as Si considered here, the
macroscopic dielectric tensor is diagonal.

\section{Computational methods}
\label{sec:methods}

For the calculation of the dielectric function, one needs 
wavefunctions $\phiik$ and eigenvalues $\eik$ at the chosen set of
$\veck$-points [See Eq. (\ref{eq:chi0})]. To generate these, we first
performed the LDA and EXX self-consistent ground-state calculations at
the experimental lattice constant, $a = 5.43 \AA$ for Si, and obtained
the KS potential $\vKS$ which serves as the only input to the TDDFT
calculation of optical spectra. The EXX calculation was carried out
using ten special-$\veck$ points, orbital kinetic energy cutoffs of
12.5 Ha, and nonlocal pseudopotentials in the separable form of
Kleinman-Bylander~\cite{Kleinman82}. We refer the reader to Ref.
\onlinecite{Stadele97} for further details of EXX calculations. In
addition, because the ALDA kernel $\Fxc^{ALDA}$ depends only on the
density, we constructed it at the end of the ground-state calculations
when fully converged densities were available.

With the converged $\vKS^{EXX}$, $\vKS^{LDA}$ and $\Fxc^{ALDA}$ in
hand, we moved on to the computation of the dielectric function and
first solved the KS equations once again at a much larger number of
$\veck$-points but with the same energy cutoff 12.5 Ha to generate
$\chi_0$. The calculation of the head and wings of $\chi_0$ needs
special care since the $\qzero$ limit of $\langle i, \veck| e^{-i\vecq
\cdot \vecr} | j,\veck+\vecq \rangle$ should be handled properly.
Applying perturbation theory one obtains the following equation
\cite{Hybertsen87},
\begin{equation}
\label{eq:q0-limit}
 \lim_{\qzero}
	\langle i, \veck| e^{-i\vecq \cdot \vecr} | j,\veck+\vecq
\rangle
	= \vecq \cdot
	\frac{ \langle i, \veck| \vecp + [V_{NL}, i \vecr] | j,\veck \rangle }
	{\epsilon_{j,\veck} - \epsilon_{i,\veck}},
\end{equation}
which implies that, due to the extra powers of the energy denominator
that appear in the head $\chi_0^{00}(\hat{\vecq};\omega)$, computation
parameters for the convergence are more or less determined by those
necessary for the head element. We use the kinetic energy cutoff 7 Ha
and 10 conduction bands for $\chi_0$ which was sufficient to reach the
convergence of $\epsilon_2(\omega)$.

For the $\veck$-point sampling, we chose to adopt a regular uniform
grid, with which one could evaluate $\lim_{\qzero} \langle i, \veck|
e^{-i \vecq \cdot \vecr} | j,\veck+\vecq \rangle$ by a finite
difference approach without resorting to the analytic replacement of
Eq. (\ref{eq:q0-limit}). In this case one had to pay the price of
performing three additional diagonalizations at $\veck+ q
\vece_{\alpha} (\alpha = x,y,z)$ for each $\veck$-point. Therefore, in
this paper we report results obtained by treating the $\qzero$ limit
according to Eq. (\ref{eq:q0-limit}) with ignoring the nonlocal
pseudopotential contributions due to the commutator $[\widehat{V}_{NL},
i \vecr]$ term. In test calculations with a small number of
$\veck$-points we found that the neglect of the nonlocal
pseudopotential contributions has little effect on the positions of
absorption peaks while the peak heights are somewhat reduced. We
slightly shift the uniform $\veck$-mesh in order to break any present
symmetry thus achieve a more effect $\veck$-point
sampling~\cite{Rohlfing98}. After checking the convergence behavior by
increasing the number of $\veck$-points and changing the shifting
vector, we used a shifted $ 17 \times 17 \times 17$ regular grid which
gives the convergence of $\epsilon_2$ up to the frequency range of at
least 6 eV. The use of a shifted uniform $\veck$-mesh results in
slightly non-identical diagonal elements of the dielectric tensor whose
differences have been also used to check the convergence with respect
to the number of $\veck$-points.

\section{Results and discussions}
\label{sec:results}

\begin{figure*}
 \begin{minipage}[H]{\linewidth}
  \centering\epsfig{file=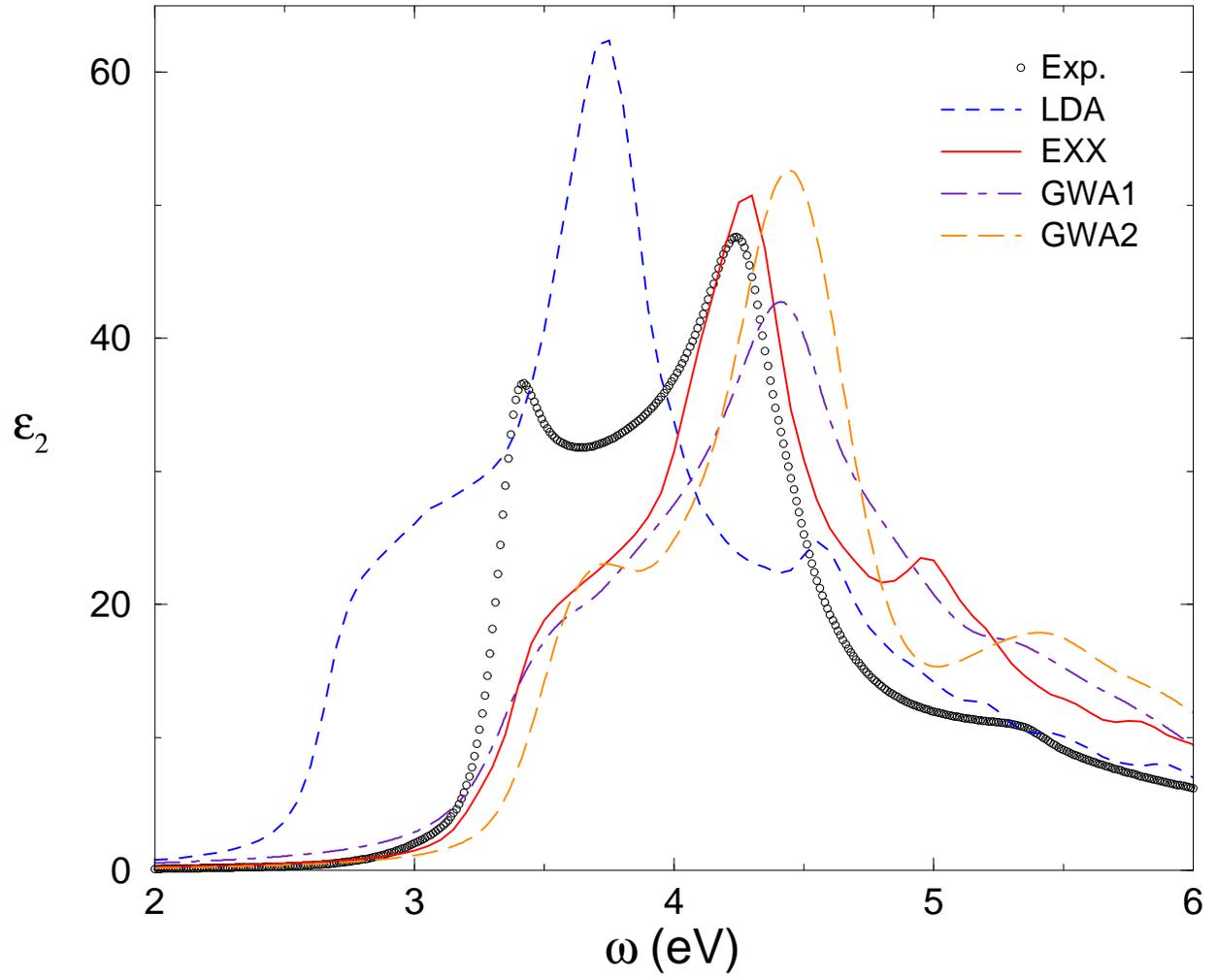}
 \end{minipage}
 \caption{Absorption spectrum of bulk silicon at the one-particle LDA
 and EXX levels. Experimental curve is from
 Ref. \protect\onlinecite{Herzinger98}. Two GWA results (GWA1 from
 Ref. \protect\onlinecite{Benedict98} and GWA2 from Ref.
 \protect\onlinecite{Albrecht98}) are also shown for comparison.}
 \label{fig:LDA-vs-EXX}
\end{figure*}

We first consider the absorption spectrum calculated at the LDA and EXX
independent particle levels.  In Fig. \ref{fig:LDA-vs-EXX}, LDA and EXX
absorption spectra are shown together with
experimental~\cite{Herzinger98} and GWA~\cite{Albrecht98,Benedict98}
curves.  Note that the two GWA results are rather different due to
operational ambiguities resulting from the post-DFT nature of the GWA
\cite{Aulbur00}, while the LDA and EXX  which strictly reside in the KS
DFT scheme do not suffer from such problems.  Within the LDA, the
absorption spectra is shifted to the lower frequency region by about 1
eV due to its well-known underestimation of bands gaps.  In addition,
the $E_1$ peak height is significantly underestimated, while that of
the $E_2$ peak is much overestimated. In contrast, the EXX absorption
spectrum shows a much better overall agreement reflecting realistic
descriptions of the band structure within the EXX~\cite{Stadele97}:
Positions of the absorption edge and the $E_2$ peak are in excellent
agreement with the experiment, and the $E_2$ peak position is even
slightly better than the GWA one. Also the height of the $E_2$ peak is
smaller than the LDA value and is much closer to the experimental data.
One noticeable disagreement is the absence of the $E_1$ peak as in the
LDA spectrum.  This, however, does not reflect the deficiency of the
EXX approach, but the excitonic nature of the $E_1$ peak
\cite{Albrecht98,Benedict98,Rohlfing98} which cannot be described at
the independent particle level as can be also seen in the GWA curves.

We feel that the excellent features of the EXX absorption spectrum
warrant more discussions. Perturbation theory along the adiabatic
connection shows that KS eigenvalue differences are well-defined
approximations of excitation energies \cite{Gorling96b}. Indeed Umrigar
{\it et al.} recently found for several atoms that quasi-exact KS
eigenvalue differences are in surprisingly good agreement with
experimental excitations~\cite{Umrigar98}. In addition, applying the
quasi-exact KS potential to the calculation of the quantum defects for
the Ne atom, Al-Sharif {\it et al.} argued that the KS potential of a
neutral atom (with $N$ electrons) corresponds to the Green-function
``potential'' of its daughter cation ($N-1$ electrons), or equivalently
the KS equation for $N$ electrons corresponds to the Dyson equation
where the reference ground state is chosen with $N-1$
electrons~\cite{Al-Sharif98}. The appealing description of excitations
by KS eigenvalue differences, in contrast to the treatment of
ionizations and electron affinities in the many-body (GWA) approach,
has not drawn much attention due to the lack of realistic \xc\
approximations which are suitable for practical calculations.
Exchange-correlation potentials derived from the LDA and other
semilocal \xc\ energy functionals do not cancel Coulomb
self-interactions, decay exponentially rather than as $-1/r$ for
localized systems, and inherently lack the term that results from the
functional derivative of the pair-correlation
function~\cite{Gritsenko94}. The self-interaction-free EXX method
corrects such problems and provides accurate KS potentials and
eigenvalues. Then, we propose that the good absorption spectrum from
such EXX method represents further evidence for the $N$-electron
excitations picture within the KS scheme.

\begin{figure*}
 \begin{minipage}[H]{\linewidth}
  \centering\epsfig{file=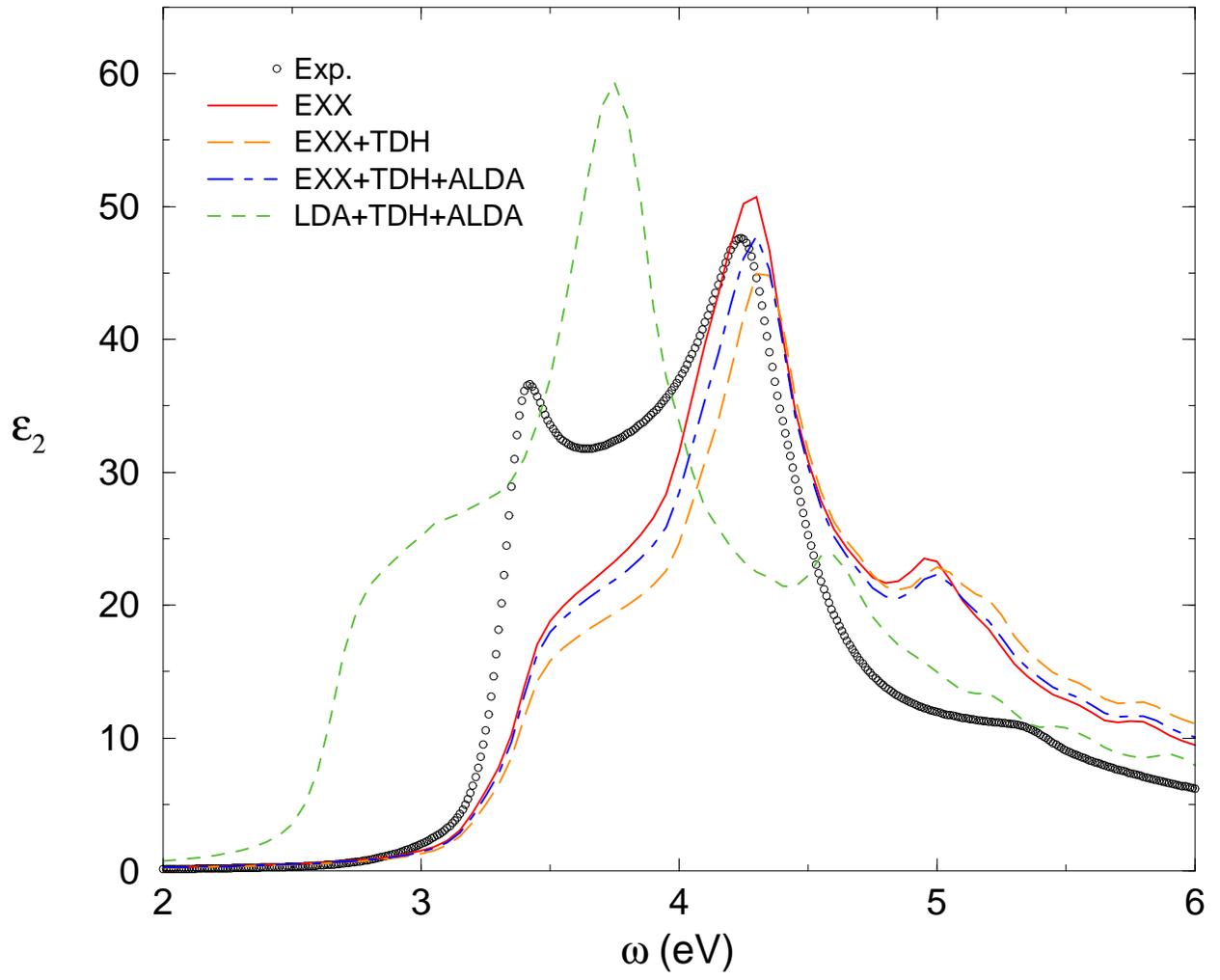}
 \end{minipage}
 \caption{Absorption spectrum of bulk silicon at the EXX, EXX+TDH,
 EXX+TDH+ALDA, and LDA+TDH+ALDA levels. Experimental data are from
 Ref. \protect\onlinecite{Herzinger98}.}
 \label{fig:EXX+TDLDA}
\end{figure*}

Now we consider the effects of the inclusion of two-particle
interactions via TDLDA on top of the independent-particle EXX result.
The Hartree-only or RPA (EXX+TDH) and full TDLDA (EXX+TDH+ALDA) spectra
are shown together with the EXX independent-particle spectrum in Fig.
\ref{fig:EXX+TDLDA}.  Granted the overall good feature of the EXX
spectrum, the most interesting question is that whether the ALDA can
generate the excitonic $E_1$ peak.  Fig. \ref{fig:EXX+TDLDA} shows that
this is not the case. TDH alone suppresses the EXX spectrum and shifts
the $E_2$ peak to a slightly higher frequency region, while adding the
ALDA counteracts such effects and more or less restores the EXX
independent-particle spectrum.  In Fig. \ref{fig:EXX+TDLDA} we also
display the TDLDA absorption spectrum based on the LDA band structure
(LDA+TDH+ALDA) to show that the LDA independent-particle result (see
Fig. \ref{fig:LDA-vs-EXX}) is again insignificantly changed by TDLDA
effects.

\section{Conclusions}
\label{sec:conclusions}

In this work, by calculating the frequency-dependent dielectric
function of bulk Si we studied the quality of EXX one-particle band
structure and the description of excitonic effects within the TDLDA. We
showed that, at the EXX independent-particle level, the position of the
$E_2$ peak is in good agreement with experimental data and is even
slightly better than that obtained in the GWA approach. We argued that
this is because the EXX approach, unlike procedures employing the LDA
and semilocal density-functionals, yields realistic KS orbitals and
eigenvalues which accurately model electronic excitations at a fixed
particle number. Employing TDDFT within the ALDA on top of the EXX band
structure however results in only a minor modification of the EXX
spectrum and fails to produce the excitonic $E_1$ peak. Given the fact
that proper inclusion of excitonic effects is mandatory for the
complete description of optical excitations, devising a TDDFT scheme
that overcomes the deficiency of the ALDA kernel will be an interesting
task. Work along this line using the nonadiabatic and nonlocal EXX
kernel is in progress~\cite{TDEXX}.

\acknowledgements 
We thank Professor W. Domcke for his support including computational
resources.  This work was supported by the Deutsche
Forschungsgemeinschaft and the Fonds der Chemischen
Industrie. Y.-H. Kim also acknowledges receipt of a Humboldt
Foundation award.


\end{document}